\documentclass{vietnam}

\def\be{\begin{equation}}
\def\ee{\end{equation}}
\def\bea{\begin{eqnarray}}
\def\eea{\end{eqnarray}}

\newcommand{\Photo}{}

\begin{document}
\vspace*{4cm}
\title{IceCube's Neutrinos: The beginning of extra-Galactic neutrino astrophysics?}

\author{E. Waxman}

\address{Department of Particle Physics and Astrophysics, Weizmann Inst. of Science\\
Rehovot, Israel}

\maketitle\abstracts{
The flux, spectrum and angular distribution of the excess neutrino signal detected by IceCube between $\sim 50\, {\rm TeV}$ and $\sim 2\, {\rm PeV}$ are inconsistent with those expected for Galactic sources. The coincidence of the excess, $E_{\nu}^{2}\Phi_{\nu}=3.6\pm1.2\times10^{-8}\, {\rm GeV/ cm^{2} sr\, s}$, with the Waxman-Bahcall (WB) bound, $E_{\nu}^{2}\Phi_{\rm WB}\approx3.4\times10^{-8}\, {\rm GeV/ cm^{2} sr\, s}$, is probably a clue to the origin of IceCube's neutrinos. The most natural explanation of this coincidence is that both the neutrino excess and the ultra-high energy, $>10^{19}$~eV, cosmic-ray (UHECR) flux are produced by the same population of cosmologically distributed sources, producing CRs, likely protons, at a similar rate, $E^2d\dot{n}/dE\approx0.5\times10^{44} \, {\rm erg/Mpc^{3}yr}$ (at $z=0$), across a wide range of energies, from $\sim 10^{15}$~eV to $> 10^{20}$~eV, and residing in environments (such as starburst galaxies) in which CRs of rigidity $E/Z< 10^{17}$~eV lose much of their energy to pion production. Identification of the neutrino sources will allow one to identify the UHECR accelerators,
to resolve open questions related to the accelerator models, and to study neutrino properties (related e.g. to flavor oscillations and coupling to gravity) with an accuracy many orders of magnitude better than is currently possible. The most promising method for identifying the sources is by association of a neutrino with an electromagnetic signal accompanying a transient event responsible for its generation. The neutrino flux that is produced within the sources, and that may thus be directly associated with transient events, may be significantly lower than the total observed neutrino flux, which may be dominated by neutrino production at the environment in which the sources reside.
}

\section{Introduction}
\label{sec:intro}

The detection of MeV neutrinos from the Sun enabled direct observations of nuclear reactions in the core of the Sun, as well as studies of fundamental neutrino properties\cite{Solar_nus}. MeV neutrino "telescopes" are capable of detecting neutrinos from supernova explosions in our local Galactic neighborhood, at distances $<100$~kpc, such as supernova 1987A. The detection of neutrinos emitted by SN1987A provided a direct observation of the core collapse process and constraints on neutrino properties\cite{SN_nus}. The main goal of the construction of high energy, $>1$~TeV, neutrino telescopes\cite{Gaisser95} is the extension of the distance accessible to neutrino astronomy to cosmological scales\cite{WnuAstro}.

The existence of extra-Galactic high-energy neutrino sources is implied by cosmic-ray observations. The cosmic-ray spectrum extends to energies $\sim10^{20}$~eV, and is likely dominated beyond $\sim10^{19}$~eV by extra-Galactic sources. The composition of the UHECRs is uncertain and their origin is unknown\cite{Lemoine2013}. Assuming that UHECRs are charged nuclei accelerated electromagnetically to high energy in astrophysical objects, some fraction of their energy is expected to be converted to high energy neutrinos through the decay of charged pions produced by the interaction of cosmic-ray protons/nuclei with ambient gas and radiation.

The upper bound derived by Waxman \& Bahcall on the intensity of extra-Galactic neutrinos\cite{WBbound} implies that km-scale (i.e. giga-ton) neutrino telescopes are required to detect the expected diffuse extra-Galactic flux in the energy range of $\sim1$~TeV to $\sim1$~PeV, and that much larger effective volume is required at higher energy\cite{WnuAstro} (see fig.~\ref{fig:WBbound}). Indeed, an extra-Galactic flux of neutrinos in the energy range of $\sim1$~TeV to $\sim1$~PeV appears to have been detected with the completion of the giga-ton IceCube detector\cite{ICdetection}. In this talk I briefly discuss the implications of this detection to our understanding of the origin of UHECRs and to the prospects of high energy neutrino astrophysics.

\section{The Waxman-Bahcall bound and IceCube's neutrino excess}
\label{sec:WB}

The observed flux and spectrum of UHECRs in the range $E>10^{19.2}$~eV is consistent with a cosmological distribution of cosmic-ray sources, producing protons at a ($z=0$) rate\cite{UHECRrate}
\begin{equation}\label{eq:uhecr}
E_p^2d\dot{n}/dE_p= 0.5\pm0.15\times10^{44} {\rm erg/Mpc^{3}yr}.
\end{equation}
This rate estimate is based on the direct measurement of UHECRs and on the well understood physics of proton-CMB interaction, and is accurate (to $\sim 30\%$) as long as the composition is dominated by protons. If the composition is dominated by heavier nuclei (up to iron), the energy generation rate at $10^{19.5}\rm eV$ would change by a factor of a few.

The UHECR composition is controversial, with air-shower data from the Fly's Eye, HiRes and Telescope Array observatories\cite{NFcomp} suggesting a proton dominated composition while the Pierre Auger Observatory\cite{AugerComp} suggesting a transition to heavy elements above $10^{19}$~eV. Due to this discrepancy, and due to the experimental and theoretical uncertainties in the relevant high energy particle interaction cross sections used for modeling the shape of the air showers, it is impossible to draw a definite conclusion regarding composition based on air-shower data at this time. It should be noted that the anisotropy signal measured at high energies, combined with the absence of this signal at low energies, is an indication for a proton dominated composition at the highest energy \cite{AnisoComp}. However, the anisotropy signal is so far identified with only a $\sim2\sigma$ confidence level\cite{Lemoine2013}. In what follows we assume that protons dominate the UHECR flux, but return to the possibility of domination by heavy nuclei at the discussion.

The energy production rate, eq.~\ref{eq:uhecr}, sets an upper bound to the neutrino intensity produced by sources, which are optically thin for high-energy nucleons to $p\gamma$ and $pp(n)$ interactions. For sources of this type, the energy generation rate of neutrinos can not exceed the energy generation rate implied by assuming that all the energy injected as high-energy protons is converted to pions (via $p\gamma$ and $pp(n)$ interactions). The resulting all-flavor upper bound is\cite{WBbound}
\begin{equation}
E_\nu^2\Phi_{\rm WB,\, all\, flavor}=3.4\times10^{-8}\frac{\xi_z}{3}\left[\frac{(E_p^2d\dot{n}_p/dE_p)_{z=0}}{0.5\times10^{44}{\rm erg/Mpc^3yr}}\right]{\rm GeV\,cm}^{-2}{\rm s}^{-1}{\rm sr}^{-1},
\label{eq:WB}
\end{equation}
where $\xi_z$ is (a dimensionless parameter) of order unity, which depends on the redshift evolution of $E_p^2d\dot{n}_p/dE_p$. The value $\xi_z=3$ is obtained for redshift evolution following that of the star-formation rate or AGN luminosity density, $\Phi(z)=(1+z)^3$ up to $z=2$ and constant at higher $z$ ($\xi_z=0.6$ for no evolution). The numerical value (3.4) given in eq.~\ref{eq:uhecr} is obtained for equal production of charged and neutral pions, as would be the case for $p\gamma$ interactions dominated by the $\Delta$ resonance\cite{WBbound}. For $p\gamma$ interactions at higher energy, or $pp(n)$ interactions, the charged to neutral pion ratio may be closer to 2:1, increasing the bound flux by $\approx30\%$.

\begin{figure}[htbp]
\centerline{\includegraphics[width=0.5\linewidth]{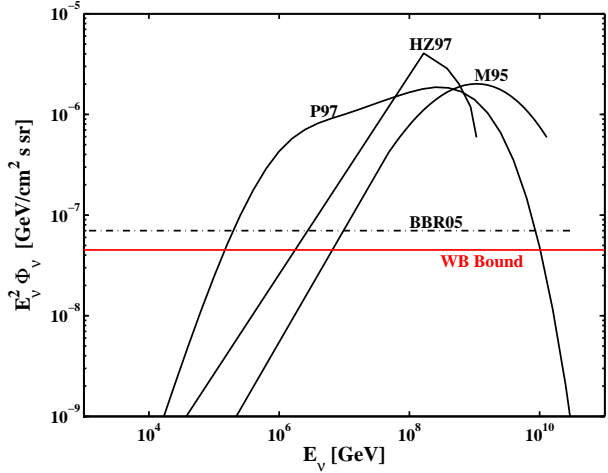}}
\caption{
The WB bound and various model predictions for the diffuse neutrino intensity produced by AGN jets. Early ("pre-bound") models [12]
predicted an intensity exceeding the bound by several orders of magnitudes, and were used as the basis for the estimate that detectors with effective mass $\ll1$~giga-ton ($\ll1 {\rm km}^2$) are sufficient for the detection of extra-Galactic astrophysical sources (see e.g. table 5 in [3]).
The fact that these models predict fluxes exceeding the bound implies, however, that they are inconsistent with UHECR observations. As the validity of the bound became widely accepted (to the point that its derivation, predictions and implications are sometimes referred to as "generic" without proper, or even any, reference, e.g. [7], [13]),
it became clear that $\ge1$~giga-ton detectors are required, and the predictions of more recent ("post-bound") models [14]
became closer to the bound. The wide range of AGN model predictions reflects the limited predictive power of the models: The predicted neutrino intensity depends strongly on the assumptions adopted. The figure is adopted from ref. [6]
and therefore presents an upper bound on the muon neutrino (and anti neutrino) intensity, neglecting oscillations and using a normalization of $E_p^2d\dot{n}_p/dE_p=1\times10^{44}{\rm erg/Mpc^3yr}$. Including oscillations, which change the $\nu_e:\nu_\mu:\nu_\tau$ flavor ratio from $1:2:0$ to $1:1:1$ [15],
and using the updated normalization, $E_p^2d\dot{n}_p/dE_p=0.5\times10^{44}{\rm erg/Mpc^3yr}$, reduces the muon neutrino upper bound by a factor of 4.
}
\label{fig:agn}
\end{figure}

Figure~\ref{fig:agn}
presents a comparison of the WB bound with various model predictions for the diffuse
neutrino intensity produced by Active Galactic Nuclei (AGN) jets. These models, which were used to estimate the detector size required for the detection of extra-Galactic astrophysical sources, typically predicted an intensity exceeding the bound by several orders of magnitudes\cite{early_agn}, suggesting that detectors with effective mass $\ll1$~giga-ton ($\ll 1{\rm km}^2$) may be sufficient: The estimated detection rate of extra-Galactic astrophysical neutrinos in a 0.1~km$^2$ detector was typically $\sim100$ events per-steradian per year (e.g. table 5 of ref. [3]). As the valisdity of the bound became widely accepted (to the point that it is considered a "generic result", which is sometimes quoted without proper or even any reference, e.g. [7,13]),
it became clear that $\ge1$~giga-ton detectors are required, and the predictions of more recent ("post-bound") models \cite{late_agn} became closer to the bound.

\begin{figure}[htbp]
\centerline{\includegraphics[width=0.6\linewidth]{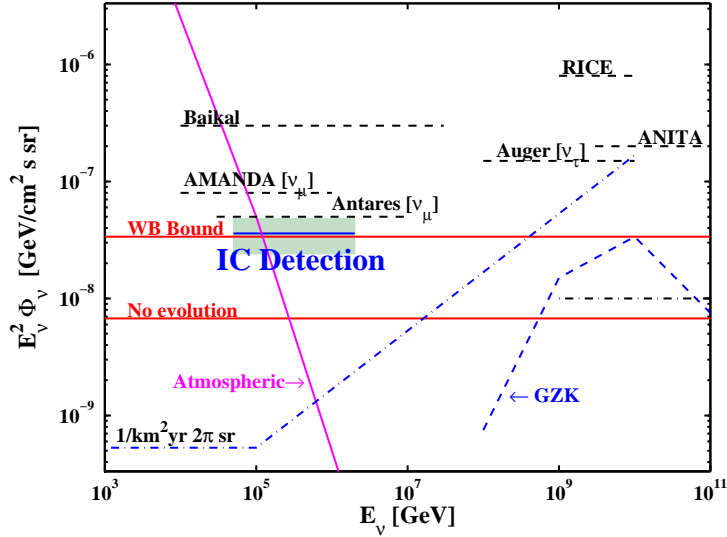}}
\caption{
The upper bound imposed by UHECR observations on the extra-Galactic (all flavor) high energy neutrino intensity (lower-curve: no evolution of the energy production rate, upper curve: assuming evolution following star formation rate), eq.~2, compared with the atmospheric muon-neutrino background, with experimental upper bounds (dashed black lines), and with the IceCube detection [7]. Shown are the muon and all flavor upper bounds of the optical Cerenkov observatories AMANDA [16], ANTARES [17] and BAIKAL [18], the all flavor upper bounds of the coherent Cerenkov radio detectors RICE [19] and ANITA [20], and the $\nu_\tau$ upper bound of the PAO [21]. The curve labelled "GZK" shows the neutrino intensity expected from UHECR proton interactions with micro-wave background photons [22]. The black dash-dotted curve is the expected sensitivity of detectors of few 100~Gton (few 100~km$^3$) effective mass (volume), that may be achieved with proposed radio detectors [23] or with proposed (optical) extensions of IceCube [24]. For a detailed discussion of the experiments see [25]. The dash-dotted blue line shows the muon neutrino intensity that would produce one neutrino induce muon in a detector with an effective area of 1~km$^2$.
}
\label{fig:WBbound}
\end{figure}

Fig.~\ref{fig:WBbound}
compares the bound with experimental upper bounds obtained by experiments preceding IceCube, and also with the recent IceCube detection. The IceCube collaboration reported the detection of 28 neutrinos in the energy range of $\sim50$~TeV to $\sim2$~PeV, which constitutes a $4\sigma$ excess above the expected atmospheric neutrino and muon backgrounds. The excess neutrino spectrum is consistent with $dn/dE_\nu \propto E_\nu^{-2}$, its angular distribution is consistent with isotropy, and its flavor ratio is consistent with $\nu_e:\nu_\mu:\nu_\tau=1:1:1$. It should be noted that the spectral shape, angular distribution and composition are currently poorly constrained due to the low statistics. The best fit normalization of the intensity is $E_\nu^2\Phi_\nu=3.6\pm1.2\times10^{-8}{\rm GeV/cm^2 s\, sr}$, coinciding in normalization and spectrum with the WB bound.

\section{Discussion}
\label{sec:discussion}

\subsection{The origin of IceCube's neutrinos}
\label{sec:origin}

The intensity associated with the neutrino excess is much higher than that expected to originate from interaction of cosmic-ray protons with interstellar gas in the Galaxy\cite{Katz13,Spector13}, and unlikely to be due to (unknown) Galactic sources, which are expected to be strongly concentrated along the galactic disk. The coincidence with the WB flux also suggests an extra-Galactic origin. We note also that a $\nu_e:\nu_\mu:\nu_\tau=1:1:1$ flavor ratio is consistent with that expected for neutrinos originating from pion decay in cosmologically distant sources, for which oscillations modify the original $1:2:0$ ratio to a $1:1:1$ ratio\cite{Learned95}.

The IceCube excess neutrinos are likely produced by interactions of high energy CR protons with protons or photons, or of high energy CR nuclei with protons, which produce pions that decay to produce neutrinos. Assuming that the neutrino excess is due to extra-Galactic sources of protons, a lower limit of $E_p^2d\dot{n}_p/dE_p\ge0.5\times10^{44}{\rm erg/Mpc^3yr}$ on the local, $z=0$, proton production rate is implied. A similar limit is obtained for heavy nuclei, since photo-disintegration does not reduce significantly the energy per nucleon\cite{Spector13}. The CR energy range corresponding to the energy range of the observed neutrinos is $\approx 1A-100A$~PeV, where $A$ is the atomic number of the CRs\cite{Spector13}. The observed neutrinos may thus be produced either by sources producing CRs at a rate $E^2d\dot{n}/dE\sim0.5\times10^{44}{\rm erg/Mpc^3yr}$ and for which CRs of rigidity $E/Z<10^{17}$~eV lose most of their energy to pion production, either within the source or at source's environment (as would be the case for sources residing in starburst galaxies\cite{LWstarbursts}), or by sources producing $E^2d\dot{n}/dE\gg0.5\times10^{44}{\rm erg/Mpc^3yr}$ and for which CRs lose only a small fraction, $f(E)\ll1$, of their energy to pion production. In the latter case, the small (and likely energy dependent) energy loss fraction should compensate the large energy production rate to reproduce the observed flux and spectrum over two decades of $\nu$ energy, and the coincidence of the observed neutrino flux and spectrum with the WB bound flux and spectrum would be a chance coincidence.

The simpler explanation, which we consider to be more likely, is that both the neutrino excess and the UHECR flux are produced by the same population of cosmologically distributed sources, producing CRs at a similar rate, $E^2d\dot{n}/dE\approx0.5\times10^{44} \, {\rm erg/Mpc^{3}yr}$ (at $z=0$), across a wide range of energies, from $\sim 10^{15}$~eV to $> 10^{20}$~eV, and for which CRs of rigidity $E/Z<10^{17}$~eV lose much of their energy to pion production\cite{correction}. Note that a $dN/dE\propto E^{-2}$ power-law spectrum of accelerated particles has been observed for both non-relativistic and relativistic shocks, and is believed to to be due to Fermi acceleration in collisionless shocks\cite{Fermi} (although a first principles understanding of the process is not yet available). The absence of neutrino detection above a few PeV, which suggests a suppression of the neutrino spectrum above this energy, may be due to efficient escape of $E/Z>10^{17}$~eV CRs from the environments in which they produce the pions, as suggested for sources residing in starburst galaxies\cite{LWstarbursts}, and need not imply a cutoff in the CR production spectrum.

As noted in \S~\ref{sec:WB}, the UHECR production rate obtained under the assumption that their composition is dominated by heavy nuclei (e.g. O, Si, Fe) differs at $10^{19.5}$~eV only by factors of a few from the rate given in eq.~\ref{eq:uhecr}, inferred assuming that the flux is dominated by protons. This is due to the fact that the energy loss distance of protons due to pion production is not very different from the energy loss of heavy nuclei due to photo-disintegration\cite{Allard12}. However, the different dependence on CR energy of the photo-disintegration energy loss distance and pion production energy loss distance implies that, under the assumption that the UHECR flux is dominated by heavy nuclei, the observed UHECR spectrum requires either a generation spectrum different than $E^2d\dot{n}/dE\propto E^0$ or an energy dependent composition\cite{Allard12} (tailored to fit the spectrum). Due to this, and to the points mentioned in \S~\ref{sec:WB}, we consider a proton dominated spectrum more likely.

\subsection{Prospects}
\label{sec:prospects}

One of the main goals of the construction of high energy neutrino telescopes is to resolve the open questions associated with the origin of UHECRs: determine their composition, identify their sources, and resolve open questions related to their acceleration and to the physics of their sources. It should be noted that the large source power required for proton acceleration to $\sim10^{20}$~eV, $L>10^{46}{\rm erg/s}$, suggests that only the most powerful known astrophysical objects, Gamma-Ray Bursts (GRBs) and the brightest AGN, are capable of such acceleration\cite{Lemoine2013}. The physics of the sources, which are probably powered by mass accretion onto black holes, and of the acceleration, which likely takes place in relativistic or mildly relativistic collisionless shocks, is poorly understood, and the related open questions are among the most interesting and important open questions of high energy astrophysics. High energy neutrino observations may allow us to resolve some of them\cite{WnuAstro}.

The coincidence, in flux and spectrum, of the neutrino excess detected by Icecube with the WB bound suggests that the sources of ICecube's neutrinos are related to the sources of UHECRs, and therefore supports our hope that high energy neutrino observations will help in resolving the open questions mentioned above. As IceCube's exposure increases with time, the measurement accuracy of the flux, spectrum, flavor content and angular distribution of the neutrino excess will improve, and will allow one to increase the confidence in (or disprove) their extra-Galactic origin. Analyzing the properties of the diffuse neutrino flux provides clues regarding the nature of the sources\cite{Cholis2013}, but is unlikely to allow one to unambiguously identify them. The most important step required for resolving the UHECR puzzle, and for studying the physics of the sources, is an electro-magnetic identification of the neutrino sources. Such identification is not very likely to be achieved simply by a factor of a few increase in exposure.

First, it should be realized that the detection of several neutrinos from a steady UHECR source is highly improbable. The probability that a $10^3$~TeV muon neutrino would produce a muon passing through the detector is $\approx3\times10^{-4}$, implying that the collection area of a 1~km$^2$ neutrino detector is effectively smaller by a factor of $\approx 10^{7}$ than that of $>10^{19}$~eV CR detectors. The fact that multiple UHECR events from a single source are not clearly detected by CR experiments implies that multiple neutrinos are unlikely to be detected unless the neutrino luminosity of the sources exceeds their UHECR luminosity by a large factor ($10^3$). Second, the degree scale resolution of the neutrino telescopes will not enable one to identify a specific source lying at a cosmological distance.

The most promising method for identifying the sources is by association of a neutrino with an electromagnetic signal accompanying a transient event responsible for its generation. Luckily, the absence of sources meeting the luminosity requirement, $L>10^{46}{\rm erg/s}$, within the proton propagation distance\cite{gzk} suggests that the sources of UHECRs are very bright transients (see also [33]).
In order to enable an association of a neutrino event with an electromagnetic event, a wide field electromagnetic transient monitoring is required. It should be noted that the neutrino flux that is produced within the sources, and that may thus be directly associated with transient events, may be significantly lower than the total observed neutrino flux, which may be dominated by neutrino production at the environment in which the sources reside. For example, if UHECRs are protons produced in GRBs, the neutrino flux expected to be produced within the sources is $\approx0.1\Phi_{\rm WB}$\cite{WnB97,Meszaros08}.

In addition to studying CR sources and acceleration, detection of high energy neutrinos in association with an electromagnetic signal may provide information on fundamental neutrino properties. Detection of neutrinos from GRBs, for example, could be used to test the simultaneity of neutrino and photon arrival to an accuracy of $\sim1$~s. It is important to emphasize here that since the background level of neutrino telescopes is very low, the detection of a single neutrino from the direction of a GRB on a time sale of months after the burst would imply an association of the neutrino with the burst and will therefore establish a time of flight delay measurement. Such a measurement will allow one to test for violations of Lorentz invariance (as expected due to quantum gravity effects\cite{WnB97,Amelino98}), and to test the weak equivalence principle, according to which photons and neutrinos should suffer the same time delay as they pass through a gravitational potential. With $1{\rm\ s}$ accuracy, a burst at $1{\rm\ Gpc}$ would reveal a fractional difference in (photon and neutrino) speed of $10^{-17}$, and a fractional difference in gravitational time delay of order $10^{-6}$ (considering the Galactic potential alone). Previous applications of these ideas to supernova 1987A, yielded much weaker upper limits: of order $10^{-8}$ and $10^{-2}$ respectively\cite{Bahcall89}. Note that at the high neutrino energies under discussion deviations of the propagation speed from that of light due to the finite mass of the neutrino lead to negligible time delay even from propagation over cosmological distances (less than $\sim10^{-10}$~s at 100~TeV).

High energy neutrinos are expected to be produced in astrophysical objects predominantly by the decay of charged pions, which lead to the production of neutrinos with flavor ratio $\nu_e:\nu_\mu:\nu_\tau=1:2:0$ (here $\nu_l$ stands for the combined flux of $\nu_l$ and $\bar\nu_l$). Neutrino oscillations then lead to an observed flux ratio on Earth of
$\nu_e:\nu_\mu:\nu_\tau=1:1:1$\cite{Learned95}. Detection of neutrino induced $\tau$'s, rather than $\mu$'s, would be a distinctive signature of such oscillations, provided the sources are understood well enough (see, e.g. [38]).
It has furthermore been pointed out that flavor measurements of astrophysical neutrinos may help determining the mixing parameters and mass hierarchy, and may possibly enable one to probe new physics\cite{Winter10}.

Finally, we note that detectors sensitive to the diffuse flux of GZK neutrinos\cite{gzk_nu} (see fig.~\ref{fig:WBbound}), expected to be produced by the interaction of UHECR protons with cosmic microwave background photons\cite{gzk}, will enable one to test the hypothesis that the UHECRs are protons of extra-Galactic origin.


\begin{thebibliography}{00}
\bibitem{correction}
  Note that, contrary to the claim following eq.~3 of ref. [13],
  this does not imply an extra-Galactic CR production rate higher than that given in eq.~\ref{eq:uhecr} (the local $10^{15}$~eV CR flux is dominated by Galactic sources and enhanced by Galactic confinement). A more detailed discussion, compared to that given in this paper, that clarifies this point at greater length may be found at [26].
\end{thebibliography}
\end{document}